\documentclass{mn2e}
\usepackage{amssymb}
\usepackage{graphicx}
\newcommand{\be}{\begin{equation}}
\newcommand{\ee}{\end{equation}}
\newcommand{\Msun}{{\rm M}_{\odot}}

\newcommand{\lum}{erg s$^{-1}$}
\newcommand{\nustar}{NuSTAR J095551+6940.8}

\begin{document}
\title[Accreting neutron star as an ULX]
{The accretion regimes of a highly magnetised NS: the unique case of \nustar}
\author[S.~Dall'Osso et al.]{Simone Dall'Osso$^1$, Rosalba Perna$^1$, Alessandro Papitto$^2$, Enrico Bozzo$^3$, 
Luigi Stella$^4$\\
$^1$ Department of Physics and Astronomy, Stony Brook University, Stony Brook, NY, USA\\
$^2$ Institute of Space Sciences (ICE, CSIC–IEEC), Carrer de Can Magrans, S/N, E-08193 Barcelona, Spain\\
$^3$ ISDC, University of Geneva, Chemin d'Ecogia 16, CH-1290 Versoix, Switzerland\\
$^4$ INAF - Osservatorio Astronomico di Roma, via di Frascati 33, 00044, Monteporzio Catone, Roma, Italy}

\maketitle
\label{firstpage}


\begin{abstract}
We analyze archival Chandra HRC observations of the ultraluminous
accreting pulsar M82-X2 (\nustar), determining an upper limit of $< 1.7\times 10^{38}$~\lum~to its luminosity at an 
epoch at which it was undetected. Combined with other recent measurements, this confirms that the source X-ray emission 
has been highly variable during the last 15 years, ranging
from a maximum of   $10^{40}$ \lum through intermediate values  $\sim$ a few $\times 10^{39}$ \lum, and down 
to a minimum that must be below the current detection threshold $\sim 2 \times 10^{38}$ \lum. We interpret these results by
  means of a magnetically-threaded disk model: when at peak luminosity, 
  the neutron star (NS) is close to spin equilibrium, its inner disk
  edge $r_m \sim 10^8$~cm is approximately half the corotation radius
  $r_{co}$, and radiation pressure dominates the disk out to $r_{tr}
  \lesssim 10^9$ cm. In the radiation pressure-dominated regime,
  $r_m$ grows very slowly as the mass inflow rate drops: as a
  result, $r_m < r_{co}$ remains valid until $\dot{M} \gtrsim$  $\dot{M}_E$, the
  Eddington accretion rate, allowing a wide range of accretion luminosities to the NS. 
  Once $\dot{M} < \dot{M}_E$ accretion onto the NS is inhibited because $r_m > r_{co}$, 
  and the source luminosity is expected to drop by a large factor.  We conclude that
  a magnetically threaded, radiation pressure-dominated disk, around a
  highly magnetized NS ($B \lesssim 10^{13}$ G) offers the best
  intepretation for all the currently observed properties of \nustar. 
  This source offers an unprecedented opportunity to study the disk-magnetosphere 
  interaction in a new regime of supercritical accretion, and across the transition between-radiation
  pressure and gas-pressure dominance inside the disk.

\end{abstract}


\section{Introduction}
\label{sec:intro}
The discovery by {\em NuSTAR} (Bachetti et al. 2014) of coherent
pulsations in the X-ray emission of an Ultra-Luminous X-ray source
(ULX) has challenged our understanding of both ULXs as well as of
accretion onto magnetized neutron stars (NSs).  The source, observed
at a peak X-ray luminosity $L_X$ of $\sim 10^{40}$~erg~s$^{-1}$
(assuming isotropic emission) has triggered numerous studies aimed at
discerning its nature and properties, and in particular the magnetic
field strength of the accreting NS in it.

The early works, based on the value of the luminosity in this
'high-state', reached however conclusions which were not univocal. In
particular, a magnetar-like magnetic field of $\sim 10^{14}$~G was
derived by Eksi et al. (2014), while Lyutikov (2014) favoured a
$B$-field $\sim 10^{13}$~G. A more 'typical' NS field of $10^{12}$~G
was suggested by the analysis of Christodoulou et al. (2014; see also
Bachetti et al. 2014), while
Kluzniak \& Lasota (2014) argued in favour of a very low field, $<
10^9$~G. A mixed magnetic field topology, with a low dipolar field and
magnetar-strength multipoles was envisaged by Tong (2015).  These
different conclusions stemmed from the combination of
different assumptions: i.e. isotropic emission versus
beamed, spin equilibrium or else purely material
accretion torques versus magnetically-threaded disks.

In Dall'Osso et al. 2015 (paper~I from here on), we analyzed the source properties
by considering a magnetically-threaded disk without prior assumptions on
how close the NS is to spin equilibrium.  Firstly, we used the
average value of the period derivative $\dot{P}$ (over 4 sets of
observations) to show that there exists a continuum of solutions in
the $B$ vs. $\dot{M}$ (or $L_X$) plane, and that
the solutions found by previous investigators correspond to particular
cases of the more general solution.  Second, we included in the
analysis the observed $\dot{P}$ variations to show that only high-field
solutions, corresponding to $B\gtrsim 10^{13}$~G, could account for
 those fluctuations (over four time intervals) without
requiring major changes in $\dot{M}$, which would be at odds with the
approximately constant X-ray emission of the source during the same
time.

All these works relied on the \nustar~observations (Bachetti
et al. 2014) of the source in its 'high' state, with
$L_X\sim10^{40}$~erg~s$^{-1}$. However, an analysis of archival {\em
Chandra} data over a period of 15 years by Brightman et al. (2015)
revealed that the source can be found in several accretion states,
with large luminosity variations, down to $L_X\sim {\rm a~few}~\times
10^{38}$~erg~s$^{-1}$.  In addition, Brightman et al.  used {\em
NuSTAR} data to study the spectrum of the pulsed component. They found
it broadly consistent with typical spectra of accreting pulsars in High Mass
X-ray Binaries (HMXBs).
These archival {\em Chandra} observations were analyzed also
by Tsygankov et al. (2015), who suggested a possible bimodal distribution of the 
luminosity, with two well-defined peaks
separated by a factor $\sim$ 40. They interpreted this bimodality
as due to transitions between the accretion and the propeller
phase in a NS with a magnetic field $\sim 10^{14}$~G. 

Here, we begin by performing an analysis of two archival {\em Chandra/HRC}
observations which were not previously published due to lack of spectral information.
We show that, in one of those observations, the source was not detected.
 Hence, over the 15 years of observations, the
source has displayed 3 states: a high-luminosity one with
$L_X\sim 10^{40}$~erg~s$^{-1}$, an intermediate-luminosity one with
$L_X$ varying between $\sim {\rm a~few} \times 10^{38}$~erg~s$^{-1}$ to a few
$\times 10^{39}$~erg~s$^{-1}$, and a low-luminosity one with an upper
limit of $L_X <  1.7 \times 10^{38}$~erg~s$^{-1}$ (see Sec. \ref{sec:observations}).  These luminosity variations,
which encompass the Eddington limit, also straddle the regime
in which the inner disk transitions from being pressure-dominated to
radiation-dominated. As such, the source \nustar\, offers an 
unprecedented opportunity to study accretion onto a magnetized NS 
across a wide range of physical conditions. Such a study is the goal of this work.

The paper is organized as follows: Sec.2 describes the analysis of
the relevant HRC {\em Chandra} data (where the source is not detected); Sec.3
summarizes the basics of the threaded disk model, both in the gas and
in the radiation-dominated regimes.  The direct application to
\nustar~ is presented in Sec.4 for a wide range of different
models. We summarize and conclude in Sec.5.

\section{Chandra archival observations: setting a limit on the lowest luminosity 
of the source}
\label{sec:observations}

Brightman et al. (2015) analyzed {\em Chandra} archival observations
covering a 15 yr period.  A few of them were discarded either due to
excessive contamination from nearby sources, or because no spectral
information was available from the two unpublished long observations
made with the HRC only. In the majority of the remaining cases, M82-X2
was detected revealing large flux variations, with the corresponding
isotropic luminosity ranging from\footnote{Different spectral models
  cause slight luminosity differences.}  $L_X \lesssim 10^{40}$ \lum~
to $L_X \gtrsim (2-3) \times 10^{38}$ \lum. For a few observations,
Brightman et al. (2015) report detections at luminosities $\sim (1-2)
\times 10^{38}$ \lum, albeit with rather large relative errors. 
In addition to using these, here we also analyze the two long
observations made with HRC in order to assess whether the source was
actually detected and, if not, set an upper limit on its luminosity.

The HRC observations under consideration were performed on 2007
January 9 (Id. 8189 with a 61.3 ks exposure, Obs.~1 hereafter),
and January 12 (Id. 8505 with an 83.2 ks exposure, Obs.~2
hereafter). In both cases {\it Chandra} observed M82 using the High
Resolution Camera optimized for timing (HRC-S Timing). In order to
eliminate mismatches in the absolute astrometry between the two
observations, we reprojected them using the HRC Chandra image 
of   the source region taken on October 28, 1999 (ObsId. 1411). We reduced and analyzed data with
the Chandra Interactive Analysis of Observations (CIAO) version 4.7,
and reprocessed event files using the \textsc{chandra$_{-}$repro}
script, with the latest calibration files included in the CALDB
v. 4.6.8.

\begin{figure*}
\centerline{
\includegraphics[width=15.3cm]{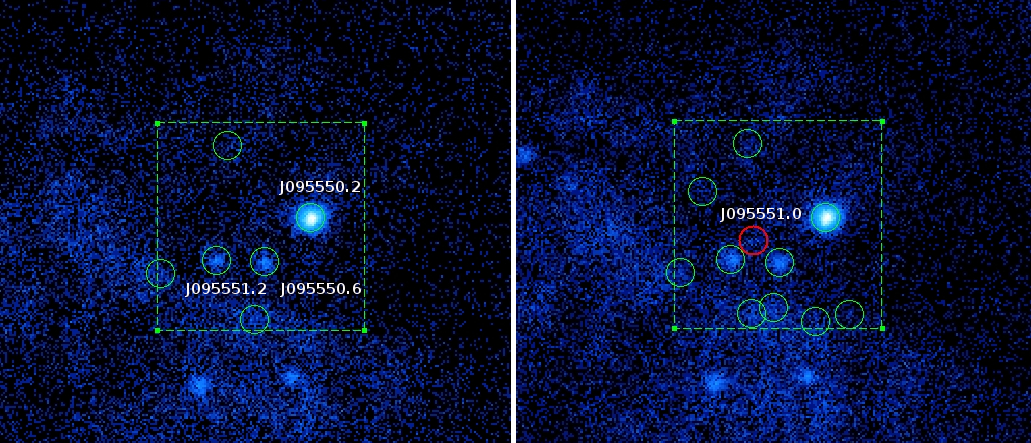}}
\caption{HRC-S image of the core of M28 obtained during Obs.~1 (left
  panel) and 2 (right panel), respectively. Green circles plotted with
  a radius of 0.9 arcsec indicate sources significantly detected by
  the wavelet algorithm used over the box indicated by the green
  dashed lines. A red circle in the right panel indicates the
  significant detection of {\nustar}.}
\vspace{-0.1in}
\label{fig:image}
\end{figure*}

We performed source detection by using the CIAO algorithm
{\textsc{wavdetect}} (Freeman et al. 2002), which correlates the {\it
  Chandra} image with wavelets of different scales and searches for
significant correlations. We searched at scales equal to 1.0,
$\sqrt{2}$, 2.0, 2$\sqrt{2}$, 4.0, $4\sqrt{2}$, 8.0, $8\sqrt{2}$, and
16.0 times the size of the pixel of HRC-S images (0.13 arcsec); some of these
oversample the point spread function of an on axis point
source (FWHM$\sim0.4$ arcsec). We set the detection threshold at 
the level expected to give at most one false source detection in the field 
that includes the brightest X-ray sources of the galaxy (green 
dashed box in Fig. \ref{fig:image}).

Fig. 1 shows the image obtained during Obs.~1 (left panel) and Obs.~2
(right panel), respectively. {\nustar} is barely detected, 
  at a significance of 3.9$\sigma$, only during Obs.~2 (see the red
circle in the right panel of Fig.~1, where the source is dubbed
J095551.0, as in Table~2 of Chiang \& Kong 2011). On the other hand,
  the source was undetected by {\textsc{wavdetect}} during
  Obs. 1.  To estimate the photon counts in the two
  observations we used the CIAO routine \textsc{srcflux}, extracting
photons from a circle with a 0.9 arcsec radius (containing 90$\%$ of
the counts expected from a point source) centered at the source
position. The background was extracted from a larger region far from the
center of the host galaxy. We obtain 0.1-10 keV count rates of
$r_1=(0.92\pm0.14)\times10^{-3}$ and $r_2=(1.55\pm0.15)\times10^{-3}$
cts s$^{-1}$ in Obs.~1 and 2, respectively.

Although the photon count (63) in Obs. 1 would lead to a detection in
the absence of background diffuse emission close to the source
position, wavelets at the considered scales did not show a significant
correlation with the spatial count distribution. Hence we ascribe most
of those counts to the diffuse emission of the galaxy. Indeed,
extracting the flux from a circular region located just north-east of
the the source location yields a count rate of
$b_1=(0.99\pm0.13)\times10^{-3}$ cts s$^{-1}$, compatible with the
flux observed from the position of M82-X2.  The diffuse emission in
the core of M82 varies by up to 70$\%$ in magnitude over spatial
scales of $\approx1.5$ arcsec, and has an extremely patchy
distribution: a reliable modelling of the diffuse emission appears
prohibitive with present instrumentation.  Therefore, since the
relative contribution of the source and the diffuse emission to the
observed flux is highly uncertain, we consider the observed values as the
most conservative upper limit on the source flux.
\begin{table}
\begin{center}
\begin{tabular}{ccc}
\hline 
\hline
Source & Obs.~1 (s$^{-1}$) & Obs.~2  (s$^{-1}$)\\
\hline
J095550.2 & $(81.6\pm1.3)\times10^{-3}$  & $(81.0\pm1.0)\times10^{-3}$  \\
J095551.0 & $(0.92\pm0.14)\times10^{-3}$ & $(1.55\pm0.15)\times10^{-3}$  \\  
J095551.2 & $(4.27\pm0.28)\times10^{-3}$ & $(4.98\pm0.26)\times10^{-3}$    \\
J095550.6 & $(4.75\pm0.31)\times10^{-3}$ & $(4.70\pm0.25)\times10^{-3}$ \\
\hline
\hline
\end{tabular}
\caption{Photon flux of the four sources in two different
  observations. J095551.0 (M82 X-2) was significantly detected only
  during Obs.~2.}
\end{center}
\end{table}
Adopting the spectral parameters of the high-luminosity state 
of M82 X-2 (a power law with
index $\Gamma=1.33$ and absorption column of $N_H=3.4\times10^{22}$
cm$^{-2}$; Brightman et al. 2015), the lowest observed count rate,
$r_1$, translates into an unabsorbed (0.1-10) keV flux of
$3.4\times10^{-13}$ erg cm$^{-2}$ s$^{-1}$ and a luminosity of
$4.4\times10^{38}$ erg s$^{-1}$ (at a distance of 3.3 Mpc; Foley et
al. 2014). Alternatively, given that the count rate from the M82-X2
region in Obs. 1 was very close to that in neighbouring source-free
regions, we also consider the possibility that $r_1$ is due to pure
background emission. In this case, with a total background of $63$
photons, we can place a $3\sigma$ upper limit of $\sim 0.38 \times
10^{-3}$ cts s$^{-1}$ to the count rate at which the source would be
statistically significant above the background.  For the same spectral
parameters as above, this translates into a (0.1-10) keV luminosity
$\simeq 1.7\times 10^{38}$ \lum~as a 3$\sigma$ upper limit during
Obs. 1 (ranging from 1.5 to 2.4 $\times 10^{38}$ \lum~for
  $\Gamma$ = 1-2).

For the detection in Obs. 2 we proceeded as follows: if $r_1$ does
indeed correspond to diffuse emission from the core of the galaxy,
then the source contribution to $r_2$ will be $r_{src} = (0.63 \pm
0.2)$ cts~s$^{-1}$.  Again, adopting the same spectral parameters as
above, $r_{src}$ translates into a (0.1-10) keV luminosity of $\sim
2.7 \times 10^{38}$ \lum~in Obs. 2, consistent with the low-luminosity
detections reported by Brightman et al. (2015).  Indeed, when combined
with those observations, our result strengthens the case for the lower
(least conservative) upper limit discussed above for Obs.~1.

Two main conclusions can be drawn from our analysis of these
additional {\em Chandra} observations, in combination with those
reported by Brightman et al. (2015). The first is that the source
is highly variable, ranging from a maximum luminosity L$_{\rm peak}
\gtrsim 10^{40}$~\lum~down to a minimum of L$_{\rm min} \sim~(2-3) \times
10^{38}$~\lum~(for a distance of 3.3 Mpc) where it can still be
detected.  Below the latter value, the background emission from the
host galaxy becomes dominant and it is extremely hard to separate it from
the emission of M82-X2.  The second conclusion is that the
source remained {\it undetected} in the first observation made with the HRC:  
we placed a very conservative upper limit 
of $4.4 \times 10^{38}$ \lum~to the source luminosity during Obs.~1 -- assuming
that the entire count rate is due to the source -- and a more 
stringent upper limit - based on a plausible assumption for the background - 
of $\sim 1.7 \times 10^{38}$ \lum. 

Whether this observed luminosity transition between Obs. 1 and 2
occurred in a continuous fashion or  in a sharp step remains to be
determined in the future, through deeper and less sparse observations
of the source.
 
\section{Basic properties of magnetically-threaded disks}
\label{sec:transition}
When accretion occurs onto a magnetic NS, the interaction
between the accreting material and the NS magnetosphere affects the
system dynamics in several ways. For disk accretion, in particular,
the main elements can be briefly summarised as follows:
\begin{itemize}
\item The accreting plasma is threaded - at least partially - by the
  stellar magnetic field. The inner disk is truncated by the resulting
  magnetic stresses at a distance $r_m$ from the NS. It is customary
  to write this truncation radius in terms of the classical Alfv\'{e}n
  radius for spherical accretion, $r_A$, with the simple expression
  $r_m \simeq \xi r_A$, where $\xi < 1$. The Alfv\'{e}n radius is
  where the magnetic pressure equals the ram pressure of the infalling
  material. It depends on the mass accretion rate, $\dot{M}$, and on
  the NS mass, radius and magnetic moment\footnote{In more detailed
    calculations (e.g., Ghosh 1996, 2006) small corrections to this
    simple functional form were derived. These have only a minor
    effect on our estimates and, for the sake of simplicity, will be
    neglected in the following.}.
For disk accretion, on the other hand,
  angular momentum balance is the most relevant constraint (see,
  e.g. Gosh \& Lamb 1979, Wang 1996). At decreasing radii, the
  magnetic stress grows faster than the viscous/material one: once it
  becomes dominant, it disrupts the disk forcing matter to fall along
  magnetic field lines.  This difference is simply encoded in the
  parameter $\xi$, the value of which is only weakly constrained: on
  observational grounds, a number of studies have indicated $\xi \sim
  (0.3-1)$ as a conservative estimate of the range of
  variation (Psaltis \& Chakrabarty 1999). Theoretically, the
  semi-analytic model by Ghosh \& Lamb 1979 (GL79 in the following),
  and recent numerical simulations (Romanova et al. 2002, 2003, 2004)
  indicate a value $\xi \approx 0.5$, with only a very weak
  dependence on $\dot{M}$, if any. In an alternative semi-analytic
  model, Erkut \& Alpar (2004) estimate values of $\xi \sim 0.35 -
  1.2$ in different sources, or even in the same source at different
  accretion stages. In an independent approach to the problem, Wang
  (1987, 1996) argued for a value of $\xi \gtrsim 1$; Bozzo et
  al. (2009) extended Wang's approach and showed
  that, in this model, $r_m$ is almost independent of the mass
  accretion rate, over a wide range of values of $\dot{M}$.  The
  relation with $r_A$ in this case is more complex, involving an
  $\dot{M}$-dependent $\xi$ (Sec.  \ref{subsec:wang}).

\item The inner edge of the disk, $r_m$, depends on the total pressure
  inside the disk, which has a contribution from both gas and 
  radiation. The relative importance of the latter
  grows at smaller radii, at a fixed mass accretion rate, as more
  energy is locally released inside the disk. In particular, for a
  given $\dot{M}$, one can define a transition radius, $r_{tr}$,
  inside which radiation pressure dominates over gas pressure (e.g., Frank et al. 2002):
\be
\label{eq:rtrans}
r_{tr} \approx 7.5 \times 10^7{\rm
  cm}\left(\frac{\alpha}{0.1}\right)^{2/21}
\left(\frac{\dot{M}}{\dot{M}_E}\right)^{16/21} \left(\frac{M}{1.4
  \Msun}\right)^{1/3} \, ,
\ee 
where $\alpha$ is the usual viscosity parameter and the Eddington mass accretion rate 
is $\dot{M}_E= 4\pi\sigma_T m_p R/c\approx 1.75 \times 10^{-8} M_{\odot}$yr$^{-1}$, with the 
corresponding Eddington luminosity $L_E = GM \dot{M}_E/R \simeq 1.75 \times 10^{38}$ \lum, for 
a 1.4 $M_{\odot}$ NS with a radius $R$ = 10 km.
As long as $r_{tr} < r_m$, only gas
pressure is important in the disk. When $r_{tr} > r_m$, on the other
hand, radiation pressure dominates the inner  disk, where
more energy is released and most of the disk-magnetosphere interaction
occurs. Following this transition, a change in the functional form of
$r_m$ occurs and, in particular, a different dependence on $\dot{M}$
will ensue\footnote{The scaling with other physics
  parameters is mostly unaffected.} (Ghosh 1996):
\begin{eqnarray}
\label{eq:radii}
r_m^{(g)} & = &  \xi \left(\frac{ \mu^4}{2GM \dot{M}^2}\right)^{1/7}~~~{\rm gas-pressure\;regime} \nonumber \\
r_m^{(r)}  & =  & A \left(\frac{\mu^4}{M \dot{M}}\right)^{1/7}~~~{\rm radiation-pressure\;regime}  \, ,
\end{eqnarray}
where the magnetic dipole moment $\mu = B_p R^3/2$ and $B_p$ is the
field strength at the magnetic pole\footnote{Note that the above
  scalings cannot be used in the Wang model: we will show in Sec.
  \ref{subsec:wang}, that this model can be described via a modified
  version of Eq. \ref{eq:radii}, with a specific $\dot{M}$-dependence
  of $\xi$ that cancels out the one in the denominator.}  .  For
consistency, the normalization, $A$, in $r^{(r)}_m$ is set so that the
two expressions are equal at the transition radius, $r_{tr}$.  As we
will show in the next section, the weaker $\dot{M}$-dependence in
radiation pressure dominated regions play an important role in
\nustar.

\item The threading magnetic field regulates the exchange of angular
  momentum between the NS and the disk. This exchange is characterized
  by two regions separated at the corotation radius, $r_{co} =
  (GM/\Omega_s^2)^{1/3}$, where the Keplerian frequency equals the NS
  spin frequency, $\Omega_s$.  In the {\it inner region} matter
  rotates faster than the NS, thus exterting a spin-up torque as it
  gives up angular momentum to the NS magnetosphere.  In the {\it
    outer region} matter rotates slower than the NS, thus causing
  spin-down as it gains angular momentum from the faster rotating
  magnetosphere.  Depending on the strength of these two terms, the NS
  can either spin-up or down while accreting; in either case, the
  system naturally evolves towards {\it spin equilibrium}, in which
  the magnetic and material torques are at equilibrium and the NS
  accretes at constant period. This occurs when $r^{(eq)}_m \approx
  (0.5-0.9)~r_{co}$, depending on poorly known microphysical details
  of the disk-magnetosphere interaction (GL79, Wang 1987, 1996, Yi et
  al. 1997, Erkut \& Alpar 2004, Parfrey et al. 2015). 
Spin equilibrium can be used to set a constraint on $B_p$, {\it if a
    transition between spin-up and spin-down is observed:} the
  luminosity at the transition yields the mass accretion rate at
  equilibrium, $\dot{M}_{eq}$, leaving only $\mu$ and $\xi$ as
    unknowns in Eq. \ref{eq:radii}.
   
\item The NS will be an {\it accretor} only if the disk extends inside
  the corotation radius, such that $r_m < r_{co}$. The accretion
  luminosity in this case is $L_{acc} \sim G M \dot{M}/R$.  If, on the
  other hand, $r_m > r_{co}$, then matter cannot
  reach smaller radii than $r_m$: in approaching the faster rotating
  magnetosphere it acquires angular momentum and is thus forced 
  outwards. This is the so-called {\it propeller regime}, in
  which accretion onto the NS is inhibited and a strong
  spin down is expected. Since matters doesn't flow inside $r_m$, only
  half of its potential energy at $r_m (\ge r_{co}$) is released, the
  other half remaining as kinetic energy of the orbital motion. The
  propeller luminosity is therefore $L_{prop} \le GM\dot{M}/(2
  r_{co})$.
\end{itemize}

\section{Luminosity variations in \nustar: model constraints}
\label{sec:constraint}

In the following we will use the observed
  luminosities of \nustar~ to set constraints on the strength of the
  NS magnetic field and on the physical regime in the accretion disk.

The luminosity range over which the source is observed, from
L$^{(ob)}_{\rm min}$ to L$^{(ob)}_{\rm peak}$, should correspond to
the {\it accretor} phase, in which $r_m \lesssim r_{co}$ and the mass
inflow proceeds all the way to the NS surface. Given the spin period
and the magnetic dipole moment of the NS, this implies the existence
of a minimum mass accretion rate, $\dot{M}_{min}$, below which $r_m$
is larger than $r_{co}$ and the NS enters the {\it propeller} regime.

The value of $\dot{M}_{min}$ and the associated luminosity,
  L$^{(th)}_{min}$, are determined by the condition $r_m \equiv
  r_{co}$. This requires that we first set the value of the
  coefficient $\xi$ in Eq. \ref{eq:radii}, and then determine the
  pressure regime in the inner parts of the disk, which will select
  the appropriate expression for $r_m$. Given the uncertainties on
  $\xi$ discussed in Sec. \ref{sec:transition}, we define here our
  ``reference" model as the one with $\xi =0.5$, close to the results
  of recent numerical simulations. This will be used for the core of
  our argument (Sec. \ref{subsec:zetamodel}), but then we will discuss
  the way our conclusions change if we let $\xi$ vary within a wider
  range, from $\lesssim 0.3$ to 1 (Sec. \ref{subsec:dependence}). We
  will also consider a specific case in which $\xi$ depends explicitly
  on $\dot{M}$.
(Sec. \ref{subsec:wang}). At the end of Sec. \ref{sec:constraint} 
we will summarize our main conclusions, showing that they 
remain robust against the use of widely different models for the disk-magnetosphere coupling.

\subsection{The ``reference model": $\xi =0.5$}
\label{subsec:zetamodel}

\begin{figure}
\centerline{
\includegraphics[width=9.5cm]{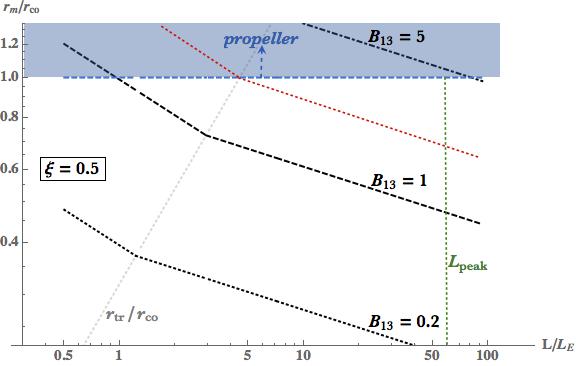}}
\caption{The ratio of the magnetospheric
  radius $r_m$ and the corotation radius $r_{co}$ as a function of
    the accretion luminosity, $L_{ac} = GM \dot{M}/R$, for three
  values of the magnetic field strength. The transition from the
  gas-pressure to the radiation-pressure dominated regime is indicated
  by the dashed grey line. The red dotted curve highlights the ``special" case  
  $B_{p,13}=2$ (cf. Sec. \ref{subsec:zetamodel}). }
\vspace{-0.1in}
\label{fig:radii}
\end{figure}

Using $\xi=0.5$ in Eq. \ref{eq:radii}, we plotted in
  Fig. \ref{fig:radii} the ratio $r_m/r_{co}$ as a function of the
  accretion luminosity, L$_{acc} = G M \dot{M}/R$, for three
  representative values of the dipole magnetic field. In the same
figure we also show the transition radius, $r_{tr}$, as a grey dashed
diagonal line: gas pressure dominates above that line, radiation
  pressure below it. Note that $r_{tr}$, hence the position of the
grey dashed line, does not depend on the value of $\xi$. 
  However, the value of the accretion rate at this transition,
  $\dot{M}_{tr}$, and of the associated luminosity, L$_{tr}$, are
  obtained from the equality $r^{(g)}_m = r_{tr}$, and thus depend
  (almost linearly) on $\xi$. In Eddington units we obtain \be
\label{eq:mdot_trans}
\frac{\dot{M}_{tr}}{\dot{M}_E}=\frac{L_{tr}}{L_E} \approx
3~B^{6/11}_{p,13} (\xi /0.5)^{21/22} \frac{R^{18/11}_6}
{M^{3/22}_{1.4}} \, ,  \ee 
where $Q_x=Q/10^x$ and 
$M_{1.4} \equiv M/(1.4M_{\odot})$.
By substituting $\dot{M}_{tr}$ in the
  condition $r^{(r)}_m = r^{(g)}_m$ one can fix the factor $A$ 
  in Eq. \ref{eq:radii}, obtaining $r^{(r)}_m = r^{(g)}_m
  (\dot{M}/\dot{M}_{tr})^{1/7}$.

Before describing in detail our results, three general conclusions can
be drawn from the previous analysis: \\ 1) If $B_{p,13} \lesssim 2$,
the condition $r_m = r_{co}$ is reached in the gas-pressure dominated
regime.  This results from L$^{(g)}_{min} < L_{tr}$, which readily
yields $B_{p, 13} < 2~(\xi/0.5)^{-7/4} M^{35/96}_{1.4} R^{-37/16}_6
(P_s/1.37)^{77/48}$.  In this regime, the minimum accretion luminosity
is \be
\label{eq:lgmin}
\frac{L^{(g)}_{min}}{L_E} \approx 0.96 B^2_{p, 13}
\left(\frac{\xi}{0.5}\right)^{7/2}
\left(\frac{P_s}{1.37}\right)^{-7/3} \frac{R^5_6} {M^{2/3}_{1.4}} \ .
\ee  2) If $B_{p,13} > 2 $, on the other hand, 
 the disk is already radiation-pressure dominated at the propeller-accretor transition.  
In this case, the minimum luminosity has an even stronger dependence on the magnetic field, scaling
like $B^4_p$, and is shifted towards larger values compared to the
extrapolation of the gas-pressure dominated case.\\ 3) If
$B_{p,13} < 2$, the flattening of the $\dot{M}$-dependence of $r_m$
beyond $\dot{M}_{tr}$ favours a much wider range of accretion
luminosities than the standard $r_m \propto \dot{M}^{-2/7}$
scaling. On the contrary, the same flattening of the
$\dot{M}$-dependence shrinks the range of accretion luminosities for
NS with $B_{p,13} > 2$, because its effect is shifting the
minimum upwards.

We can now look more closely at the behaviour of the solutions for
three representative values of the magnetic field:
\begin{enumerate}
\item For $B_{p,13} =0.2$ the disk is radiation pressure-dominated at
  all accretion rates $\ge 1.1~\dot{M}_{E}$, with the inner disk
  extending well inside the corotation radius.  The observed
  luminosity range, $\sim (1.1-60)$ L$_E$, corresponds to $r_m$ being
  between $\approx 0.22~r_{co}$ and $\approx 0.4~r_{co}$, always in
  the radiation-pressure dominated regime. Very close to the observed
  minimum, $L^{(ob)}_{min} \simeq 1.1 L_E \approx 2 \times 10^{38}$
  \lum, gas pressure takes over while the NS remains an accretor for
  further decreasing values of $\dot{M}$, until the transition to
  propeller coccurs at $L^{(th)}_{min} \approx 0.044~ {\rm L}_{E}
  (B_{p,12}/2)^2 (\xi/0.5)^{7/2} ({\rm P}_{s}/1.37{\rm s})^{-7/3}$.
  Due to the background emission from the host galaxy and the
  neighbouring sources, it will be observationally very challenging to
  track the source at these low luminosities; however, no sharp
  transition below the currently observed minimum is expected in this
  case.\\ At the same time, the high peak luminosity of the source
  would require a significant beaming if B~$\sim 10^{12}$~G, since in
  this case there is no obvious way to exceed the Eddington limit by a
  large factor. In fact, a beaming $\sim 0.15$ would also reconcile
  the measured $\dot{P} \simeq 2 \times 10^{-10}$ with the
  (beaming-corrected) luminosity. Explaining such a beaming {\it of
    the pulsed emission} represents a problem on its own (see
  discussion in paper~I).  In addition, this scenario encounters an
  even greater difficulty in the large $\dot{P}$-fluctuations measured
  while the source was near its peak emission (Bachetti et
  al. 2014). This strongly suggest that the NS was close to
  spin equilibrium at that epoch (paper I), while the result $r_m
  \simeq 0.22~r_{co}$ at the peak is hardly consistent with spin
  equilibrium, being largely below the range $r^{(eq)} \approx
  (0.5-0.95)~r_{co}$ provided by all existing theoretical
  estimates. Therefore, a new explanation would also be required for
  the large $\dot{P}$ fluctuations.

\item For $B_{p,13}=1$, radiation pressure dominates the inner disk at L$_X \ge
  3~{\rm L}_{E}$, which corresponds to $r_m \le 0.8~ r_{co}$; as
  expected, the disk is gas-pressure dominated at the minimum
  luminosity, L$^{(g)}_{min} \approx 1.7 \times 10^{38}$
  \lum~(cf. Eq. \ref{eq:lgmin}), similar to the currently observed
  minimum luminosity of M82-X2.

It is therefore a natural prediction that a spinning NS with P$_s =1.37$~s would be
detectable as an accreting source down to $\sim 2 \times10^{38}$ \lum 
{\it if it had a magnetic dipole B~$\sim 10^{13}$ G}. 

Once in propeller, the luminosity will be much lower since now the
accretion radius will be $r_{co}$ instead of $R$ (e.g. Corbet 1996). In the case of an
``ideal" (i.e. sharp) transition, the source luminosity should drop suddenly to
$\sim 5\times 10^{35}$ \lum: even in a more realistic and smoother
transition, a significant drop in luminosity must be expected just
below the currently observed minimum (e.g., Romanova et al. 2003).  More generally, an
observational determination of L$_{min}$ would allow a robust estimate
of the magnetic field strength.  In the observations discussed in Sec.
\ref{sec:observations} the source was undetected, with a 3$\sigma$
upper limit on its luminosity of $\approx 10^{38}$~\lum, consistent
with L$^{(g)}_{min}$ derived above: this limit can only be improved
with future X-ray detectors having at least the same angular
resolution of {\em Chandra} and a much larger collecting area (to
achieve the required limit within reasonable exposure times), given
the high and highly patchy background in the source region.

In addition, when B~$\sim 10^{13}$ G the Eddington limit is 
  significantly modified by the magnetically-reduced electron
scattering cross-section, which gives a maximum luminosity
$L^{(th)}_{peak} \sim 10^{40} B^{4/3}_{13}$\lum~(Paczynski 1992).
This value matches well the observed peak luminosity of
\nustar~(cf. paper~I) that, as shown in Fig. \ref{fig:radii}, is naturally
reached when $r_m \gtrsim 0.5 r_{co}$, a condition that is well
  consistent with spin equilibrium.

\item For $B_p \ge 5\times 10^{13}$~G, the equality $r^{(r)}_m =
  r_{co}$ implies that the system is in the propeller regime
  at all accretion rates lower than $\sim 85~ \dot{M}_{E}
  (B_{p,13}/5)^4$.  This limit, which is $\gtrsim$ 3 times higher than what
  would have been obtained using $r_m^{(g)}$, is a direct consequence
  of the transition to the radiation-pressure dominated regime above $\dot{M} \sim
  7.5~ \dot{M}_{E}$.

This ``high-field case" is the most problematic in light of current
observations. The NS can only enter the accretor regime at a minimum
luminosity L$^{(th)}_{min}\approx 1.7 \times 10^{40}$~\lum, as a
consequence of radiation pressure dominating the inner disk, thus violating two
constraints at a time: on the one hand, L$_{min}^{(th)}$ exceeds the
observed peak luminosity of the source and, on the other hand, all
intermediate states in which the source has a luminosity $\sim~{\rm a
  \;few}~\times 10^{39}$ \lum~would be impossible to account for,
given that L$_{prop} \sim 6\times 10^{37}$~\lum.  A stronger field
would only exacerbate this problem, given that L$^{(th)}_{min}$ scales
with $B^4_p$ in the radiation-pressure dominated regime. Note that, not accounting for
the effects of radiation pressure would allow the NS to enter the
accretor regime at a significantly lower luminosity, as was recently
found by Tsygankov et al. (2015).
  
 \end{enumerate} 
To summarize, we have shown that a NS
with B$~\sim10^{13}$ G is sufficient to interpret 
the observational properties of \nustar~ in a
straightforward fashion: its peak luminosity, the minimum
luminosity at which it is observed and the non-detections below that
value.  In addition, it was already shown that this case can also
explain the measured value of $\dot{P}$ and its fluctuations around
the peak luminosity (see Dall'Osso et al. 2015).\\ A lower field
strength might not be  excluded - although significantly lower
  $B$-fields require a high degree of fine-tuning (see paper I). A $B$-field
significantly larger than $10^{13}$ G would not match the luminosity
range over which the source is observed and can be ruled out.

\subsection{Constant $\xi$ models: dependence on $\xi$-values}
\label{subsec:dependence}

\begin{figure}
\centerline{
\includegraphics[width=9.5cm]{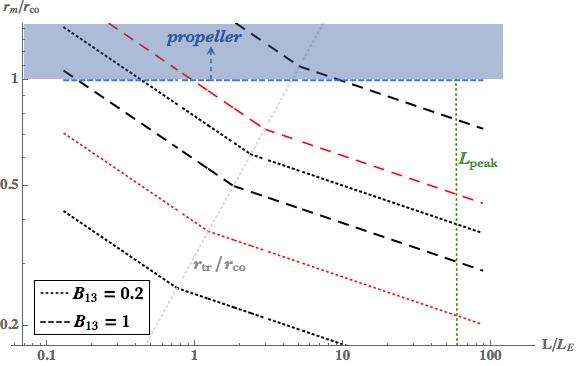}}
\caption{Same as in Fig.~\ref{fig:radii}, but 
for different values of the
  parameter $\xi$. For each value of the magnetic field, the lower
  (dotted) black curve is obtained for $\xi =0.3$ and the upper one (dashed) for
  $\xi=1$. The red curves show the results of Fig. \ref{fig:radii},
  obtained with the reference value $\xi=0.5$.}
\vspace{-0.1in}
\label{fig:rangeofradii}
\end{figure}

We next explore the dependence of our results on the parameter
$\xi$. Given the linear dependence of $r_m$ on $\xi$, for the same
magnetic field strength, a larger $\xi$ will push the system in
the propeller phase at larger values of $\dot{M}$.  Correspondingly,
the minimum luminosity at which the system can be in the
accretion phase will be larger for larger $\xi$, given a fixed
$B$-field strength. This is clear from the dependence of
  L$^{(th)}_{min}$ on $\xi$ in Eq. \ref{eq:lgmin}, and is shown in
Fig. \ref{fig:rangeofradii} for the two limiting cases $\xi = 0.3$ and $\xi=1$.
 The dependence of $\dot{M}_{tr}$ on $\xi$ (Eq. \ref{eq:mdot_trans}) is due to 
 the shifting position of the $r_m$ curves.  

Conversely, we can rewrite Eq. (\ref{eq:lgmin}) as a relation between
the magnetic dipole of the source and the value of $\xi$, if we
  know the minimum luminosity at which the NS can accrete. Recalling that
  Eq. \ref{eq:lgmin} holds for $B_{p,13} \lesssim 2$, we
  obtain \be
\label{eq:B-xi-relation}
B_{p,13} \simeq \left[\frac{L^{(g)}_{min}}{L_E}\right]^{1/2}
\left(\frac{\xi}{0.5}\right)^{-7/4} \left(\frac{P_s}{1.37
  s}\right)^{7/6} \frac{M^{1/3}_{1.4}}{R^{5/2}_6} \, .  \ee If we now
assume for $L^{(g)}_{min}$ the minimum luminosity $L^{(ob)}_{min}
\approx 10^{38}$\lum~ determined with current observations
(Sec. \ref{sec:observations}), then from Eq. \ref{eq:B-xi-relation} we
obtain $B_{13} \sim (0.4-3)$ for $\xi$ in the ($1-0.3$) range, very
similar to the ``high-B case" discussed in paper~I. These numbers can
be easily adjusted if future observations will be able to determine
the minimum luminosity with better accuracy: note, however, that the
estimated magnetic field strength scales only with the square root of
the minimum accretion luminosity.

\subsection{Wang model: a specific prescription for $\xi(\dot{M})$} 
\label{subsec:wang}
Wang (1987, 1995, 1996) derived the inner disk radius from the
condition $r^2_m B_z(r_m) B_{\phi} (r_m) = \dot{M} \left[d/dr\left(r^2
  \Omega_K\right)\right]_{r_m}$, {\it i.e.} that the magnetic stress
is equal to the material stress in the disk\footnote{The same
  condition is also considered in the GL model. However, in this
  model the radial rotation profile $\Omega(r)$ is allowed to deviate
  from keplerian in a narrow boundary layer, before reaching the inner
  disk edge in corotation with the magnetosphere. This different 
  treatment results in a different location of the inner disk edge in the GL model
  and a more pronounced dependence of $r_m$ on the
  mass accretion rate.}. Adopting different phenomenological
prescriptions for the growth and dissipation of the toroidal field in
the disk, the radial profile of $B_{\phi} (r)$ can be calculated and,
hence, the value of $r_m$ as a function of the NS magnetic field, the
mass accretion rate and several microphysics parameters that are
poorly constrained from both theory and observations.

This model predicts that, at spin equilibrium, the inner disk radius
is close to the corotation radius ($r_m^{(eq)} \gtrsim 0.9 r_{co}$),
nearly independent of the phenomenological prescriptions adopted for
the toroidal field in the disk (Wang 1995).  Bozzo et al. (2009)
generalized this approach, deriving expressions for $x =r_m / r_{co}$
in Wang-type models with various prescriptions for the mechanisms by
which the toroidal field is damped.  Here we consider two cases, which
bracket the (small) range of possible variations: {\it a)} the winding
of the field lines threading the disk is limited by magnetic
reconnection taking place in the magnetosphere; {\it b)}
the amplification of the toroidal field is damped by diffusive decay
due to turbulent mixing within the disk.
\begin{eqnarray}
\label{eq:bozzo}~
x^{-7/2}-x^{-2} & = & 4.17 \times 10^{-3} ~\frac{M^{5/3}_{1.4}P^{7/3}\dot{M}_{16}}{\gamma_{max} ~\eta^2 \mu^2_{30}}~~~~~~~a) \nonumber \\
x^{-7/2}-x^{-2} & = & 4.17 \times 10^{-3}~ \frac{\alpha}{\gamma} ~\frac{M^{5/3}_{1.4}P^{7/3}\dot{M}_{16}}{\eta^2 \mu^2_{30}}~~~b) ~\, .
\end{eqnarray}
In the above, $\alpha$ is the usual viscosity coefficient,
$\gamma=B_{\phi}/B_z$ is the magnetic pitch angle in the disk and
$\gamma_{max}$ its maximum value compatible with magnetic reconnection
out of the disk plane, and $\eta$ the factor by which the NS magnetic
field is screened by electric currents flowing in the disk. These
parameters are subject to significant uncertainties, that make it
difficult to draw quantitative conclusions. However, such
uncertainties can be bypassed {\it if a transition between spin-up and
  spin-down is observed} (Bozzo et al. 2009), since the transition
signals the point of spin equilibrium.  When expressed in terms of the
equilibrium quantities, $\dot{M}_{eq}$ (observed) and $x^{(eq)}$
(fixed by the model), Eqs. \ref{eq:bozzo} give $x$ as a function of
$\dot{M}$ and show that $r_m$ remains $\lesssim r_{co}$, nearly
constant for $\dot{M} < \dot{M}_{eq}$. In particular, accretion onto
the NS continues even at the smallest values of $\dot{M}$, and is
never quenched by the propeller. This apparently unphysical behaviour
makes it impossible, within the framework of Wang-type models, to
constrain the NS properties from luminosity variations.  However, we
can use Eqs. \ref{eq:bozzo} to estimate the NS magnetic field based on
the assumption that a spinup-spindown transition, or a very close
approach to spin equilibrium, was observed in \nustar.  This
assumption is motivated by the large torque fluctuations measured by
Bachetti et al. (2014), when the source was close to its peak
luminosity (cf. paper I). We thus set\footnote{In paper I it was shown that the spinup-spindown transition 
is likely to occur at luminosities somewhat smaller than the peak. Hence, we set for reference this 
transition at half the peak, {\it i.e.} at 30 $\dot{M}_E$. Note the weak dependence of the estimates in
Eq. \ref{eq:Bwang} on the exact value of this parameter.}
 $\dot{M}_{eq} \approx 30
\dot{M}_E$, and use the value $x^{(eq)} \simeq 0.967(0.915)$ for case
$a$($b$) in Eqs. \ref{eq:bozzo} to obtain:
\begin{eqnarray}
\label{eq:Bwang}
B_{p,13} & \approx & \frac{4.2}{\eta \gamma_{max}^{1/2}} \left(\frac{\dot{M}_{eq}}{30\dot{M}_E}\right)^{1/2} P_{1.37}^{7/6}~ \frac{M^{5/6}_{1.4}}{R^{3}_6}~~~~~~~~~~~a) \nonumber \\
B_{p,13} & \approx & \frac{0.9~\alpha_{0.1}^{1/2}}{\eta \gamma^{1/2}}  \left(\frac{\dot{M}_{eq}}{30\dot{M}_E}\right)^{1/2}P_{1.37}^{7/6} ~\frac{M^{5/6}_{1.4}}{R^{3}_6}~~~~~~~~b) \, .
\end{eqnarray}
These results are again in overall agreement with those of the
previous sections and of paper I: as a general conclusion, the
estimated $\sim 10^{13}$~G $B$-field appears robust against a variety
of independent arguments and adopted models. Note that, since $\eta <
1$ and $\gamma \gtrsim 1$, slightly larger values of the magnetic
field may possibly be favoured by Wang-type models.

Using the magnetic fields of Eqs. \ref{eq:Bwang} we can calculate
$r_A$ and, since $r_m \lesssim r_{co}$ when $\dot{M} \lesssim
\dot{M}_{eq}$, we obtain $r_m/r_A \sim (0.25-0.55)$ in \nustar, for
$\dot{M} = (\dot{M}_E - \dot{M}_{eq})$. Eqs. \ref{eq:bozzo} also imply
that $r_m$ starts decreasing approximately like $r_A$ if $\dot{M} >
\dot{M}_{eq}$ (Bozzo et al. 2009): therefore, $\xi$ remains $\gtrsim
0.55$ when $\dot{M}$ is above the equilibrium value. As a conclusion,
Wang's model may be cast in the form $r_m = \xi(\dot{M}) r_A$, where
the numerical value of the function $\hat{\xi}(\dot{M})$ ranges in the
lower half of the interval considered in the previous sections. This
is consistent with our discussion of Sec.  \ref{subsec:dependence},
that lower values of $\xi$ tend to favour larger values of the NS
magnetic field.

\section{Summary and discussion}
\label{sec:discussion}
The wide spread of luminosities of \nustar~(M82-X2)\,, encompassing a
range from highly super-Eddington, to $\sim$ Eddington to no
detection, has allowed to investigate various physically interesting
accretion regimes for a magnetized NS.

Here, we have extended the analysis of archival {\em Chandra} data
with the inclusion of two {\em HRC} observations. The source is
detected in one of them, with an estimated luminosity $\sim 2.6\times
10^{38}$~erg~s$^{-1}$. On the other hand, the source is {\em
  undetected} in the first HRC observation, during which it is
indistinguishable from the diffuse emission of the host galaxy. This
non-detection yields a conservative upper limit of $L^{(a)}_{u.l.} \ga
4.4\times 10^{38}$~erg~s$^{-1}$ to the source luminosity, assuming
that the background contribution to the counts from the M82-X2 region
is negligible.  On the other hand, by assuming that those counts are
mostly due to the background - as suggested by the count rate being
the same as in nearby background regions - we placed a more
stringent upper limit of $L^{(b)}_{u.l.} \simeq 1.7 \times 10^{38}$
\lum~to the source luminosity. 
 
We performed an in-depth analysis of the large luminosity
variations of \nustar~(M82-X2), within a magnetically-threaded disk
model, including the transition of the inner disk between gas-pressure
and radiation-pressure dominance across the various accretion states.
We explored a range of models characterized by different
prescriptions for the location of the disk truncation radius, and
found that the NS magnetic field should be in the (0.4-3)$\times
10^{13}$~G range, with a favored value of $\sim 10^{13}$~G, if accretion
onto the NS becomes inhibited below a luminosity $\simeq L_{E}$. The
latter value is fully compatible with the currently determined minimum
luminosity of the source, and with the upper limit $L^{(b)}_{u.l.}$
discussed above. The estimated $B$-field is consistent with the
conclusions of paper~I, which were based on independent
arguments.  

In our analysis, the estimated magnetic field has a degeneracy with
the poorly constrained value of $\xi$, the ratio between the disk
truncation radius and the Alfv\'{e}n radius. An independent
determination of the NS $B$-field might in principle break this
degeneracy, allowing to place interesting constraints on $\xi$, hence
on the viability of different models for the disk-magnetosphere
coupling. In accreting X-ray pulsars, one possible way to estimate the
NS magnetic field is via a high-energy break in the emission spectrum,
which is known to correlate with the cyclotron energy. However,  the cyclotron 
resonance shifts to progressively lower energies in several
bright X-ray pulsars, as the accretion luminosity grows and approaches
the Eddington limit.  This shift is attributed to the increasing
height of the accretion column above the NS surface, which therefore
samples regions of progressively decreasing $B$-field strength (Becker et al. 2012, and references
therein).  In \nustar, a spectral break at $\sim 14$ keV was detected when the
source was close to its peak emission (Brightman et al. 2015): taken
at face value, this is consistent with the break observed in several
accreting NS with $B\sim 10^{12}$ G. However, we note that \nustar~is
an extremely bright X-ray pulsar in an unprecedented accretion
regime. If this trend, and the correlation between cyclotron energy and spectral break, were 
to continue into the super-Eddington regime, then $B_p \sim 10^{13}$ G may well be
consistent with the finding of Brightman et al. (2015). In fact,
revealing a shift in the spectral break energy of \nustar~when the
source is at different luminosities would provide significant support
to this idea.

\section*{Acknowledgements}
SD was partially supported by an SFB/Transregio 7, funded by the Deutsche Forschungsgemeinschaft (DFG).\\
RP and SD acknowledge partial support by Chandra grant (awarded by SAO) ARS-16005X.

\end{document}